\documentclass{aastex62}
\usepackage{geometry}               
\usepackage{graphicx}
\usepackage{amssymb}
\usepackage{epstopdf}
\usepackage{fullpage}
\usepackage{natbib}
\usepackage{subfigure}
\usepackage{epsfig}
\usepackage{multirow}
\usepackage[figuresright]{rotating}
\usepackage{amsmath}

\def\aj{AJ}                   % Astronomical Journal
\def\apj{ApJ}                 % Astrophysical Journal
\def\apjl{ApJL}                % Astrophysical Journal, Letters
\def\apjs{ApJS}               % Astrophysical Journal, Supplement
\def\ao{Appl.~Opt.}           % Applied Optics
\def\aap{A\&A}                % Astronomy and Astrophysics
\def\icarus{Icarus}           % Icarus

\def\mnras{MNRAS}             % Monthly Notices of the RAS
\def\ssr{Space~Sci.~Rev.}     % Space Science Reviews
\def\nat{Nature}              % Nature
\def\jgr{J.~Geophys.~Res.}    % Journal of Geophysics Research
\def\jqsrt{J.~Quant.~Spec.~Radiat.~Transf.}
\def\procspie{Proc.~SPIE}   % Proceedings of the SPIE
\def\jgra{J. Geophys. Res-Atmos}     %Journal of Geophysical Research-Atmospheres
\shorttitle{Hydrohalite on M-dwarf Planets}
\shortauthors{Shields et al.}

\begin{document}
\title{Hydrohalite Salt-albedo Feedback Could Cool M-dwarf Planets}

\correspondingauthor{Aomawa Shields}
\email{shields@uci.edu}

\author{Aomawa L. Shields}
\affil{Department of Physics and Astronomy \\
University of California, Irvine \\
4129 Frederick Reines Hall \\
Irvine, CA\\ 
92697-4575 USA}

\author{Regina C. Carns}
\affiliation{Polar Science Center \\
Applied Physics Laboratory\\
University of Washington\\
Seattle, WA\\
98105-6698 USA}

%\author{Aomawa L. Shields\altaffilmark{1}
%and Regina C. Carns\altaffilmark{2}\\
%\vspace{5mm}
%\textbf{Draft version date}: \today\\
%}

%\altaffiltext{1}{\bf{Corresponding author: Aomawa Shields, Department of Physics and Astronomy, University of California, Irvine, 4129 Frederick Reines Hall, Irvine, CA 92697-4575. shields@uci.edu.}}
%\altaffiltext{2}{Polar Science Center, Applied Physics Laboratory, University of Washington, Seattle, Washington, USA}

\begin{abstract}
A possible surface type that may form in the environments of M-dwarf planets is sodium chloride dihydrate, or ``hydrohalite" (NaCl $\cdot$ 2H$_2$O), which can precipitate in bare sea ice at low temperatures. Unlike salt-free water ice, hydrohalite is highly reflective in the near-infrared, where M-dwarf stars emit strongly, making the effect of the interaction between hydrohalite and the M-dwarf SED necessary to quantify. We carried out the first exploration of the climatic effect of hydrohalite-induced salt-albedo feedback on extrasolar planets, using a three-dimensional global climate model. Under fixed CO$_2$ conditions, rapidly-rotating habitable-zone M-dwarf planets receiving 65\% or less of the modern solar constant from their host stars exhibit cooler temperatures when an albedo parameterization for hydrohalite is included in climate simulations, compared to simulations without such a parameterization. Differences in global mean surface temperature with and without this parameterization increase as the instellation is lowered, which may increase CO$_2$ build-up requirements for habitable conditions on planets with active carbon cycles. Synchronously-rotating habitable-zone M-dwarf planets appear susceptible to salt-albedo feedback at higher levels of instellation (90\% or less of the modern solar constant) than planets with Earth-like rotation periods, due to their cooler minimum day-side temperatures. These instellation levels where hydrohalite seems most relevant correspond to several recently-discovered potentially habitable M-dwarf planets, including Proxima Centauri b, TRAPPIST-1e, and LHS 1140b, making an albedo parameterization for hydrohalite of immediate importance in future climate simulations.  
\end{abstract}
%\maketitle
\pagebreak

\section{Introduction} \label{sec:intro}
Planetary climate and habitability are strongly affected by the interaction between the spectral energy distribution (SED) of the planet's host star and the planet's unique surface composition. Surface types exhibit wavelength-dependent radiative properties, making their interactions with different host star SEDs distinct and challenging to model. This complexity is particularly acute for planets orbiting M-dwarf stars, whose near-infrared (near-IR) emission coincides with clear differences in the albedo of specific surface types such as ocean, water ice, and snow at these wavelengths \citep{Joshi2012, Shields2013, Shields2014, Shields2016b}. 

Here we consider the radiative properties and impact on habitability (the capability of sustaining liquid water on some part of a planet's surface, e.g. \citealp{Seager2013}) of a surface type never before considered in the context of extrasolar planets\textemdash sodium chloride dihydrate, or ``hydrohalite" (NaCl $\cdot$ 2H$_2$O). At low temperatures (T$<$ -23$^\circ$C) on planets with oceans, salt within brine inclusions in bare sea ice can precipitate in crystal form, eventually forming a hydrohalite crust \citep{Light2016, Carns2016}. As shown in Figure 1, hydrohalite is much more reflective than bare sea ice in the near-IR, resulting in higher broadband albedos than even snow. The net impact of the surface salt-albedo feedback mechanism generated by hydrohalite formation on extrasolar planets has not previously been explored.

\begin{figure}[!htb]
\begin{center}
 \includegraphics[scale=0.70]{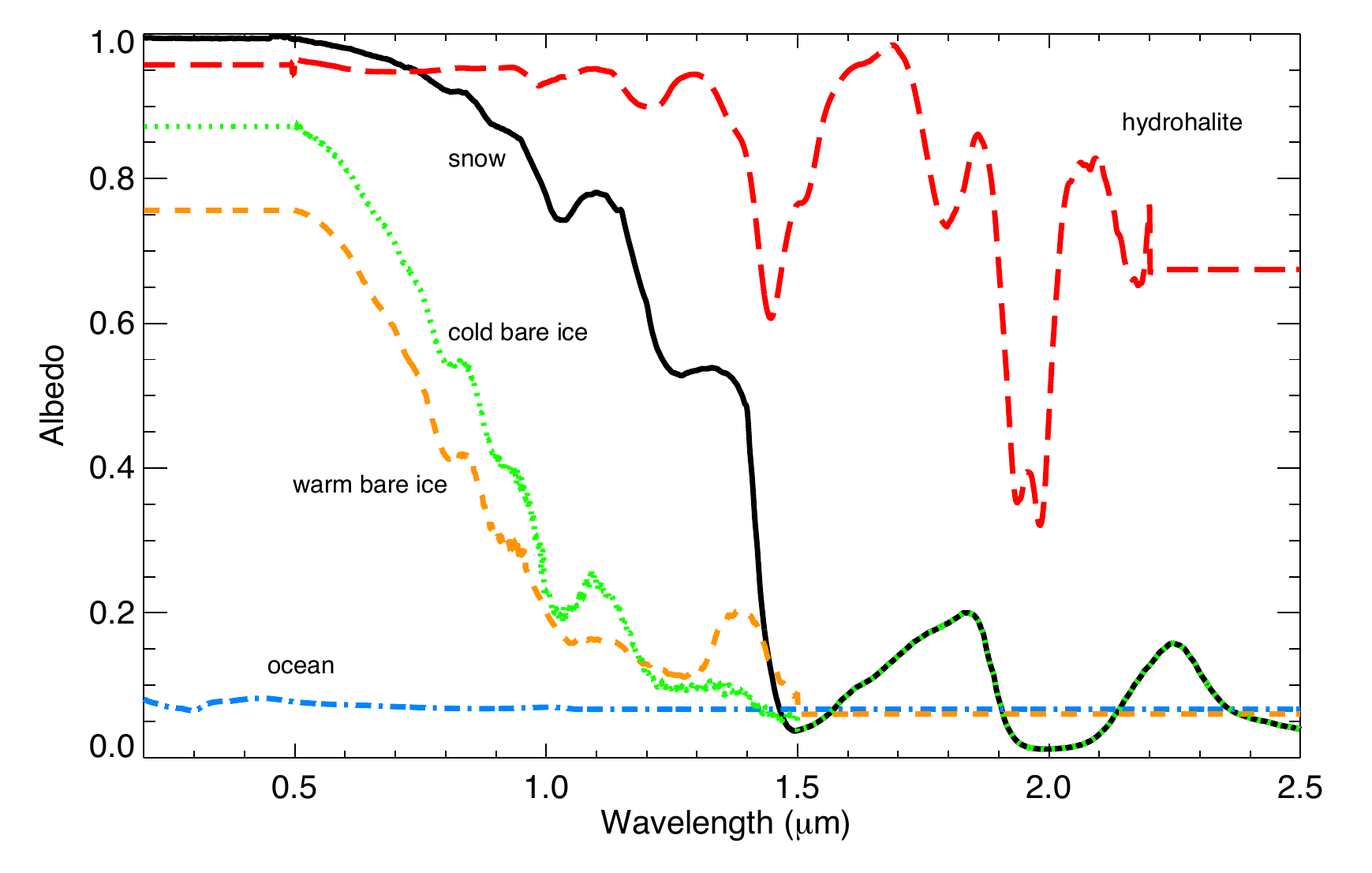}
\caption{The spectral distribution of sodium chloride dihydrate, or ``hydrohalite" (NaCl $\cdot$ 2H$_2$O) from Light \emph{et al.} (\citeyear{Light2016}), fine-grained snow (H$_2$O, free of salts, \citealp{Hudson2006}), cold and warm bare ice \citep{Light2016}, and ocean, from Brandt \emph{et al.} (\citeyear{Brandt2005}).} 
\label{Figure 1.}
%\end{minipage}
\end{center}
\end{figure}

%The anticipated climatic differences of hydrohalite crusts could therefore be significant for M-dwarf planets compared with G-dwarf planets, whose host stars emit far less radiation at near-IR wavelengths.
Studies of hydrohalite crusts have been applied to episodes of global-scale glaciation during the Neoproterozoic periods (600-800 million years ago) on the Earth \citep{Light2009, Light2016, Carns2016}, so-called ``Snowball Earth" events \citep{Kirschvink1992}.  Low surface temperatures ($< 30^\circ$C) in the tropics during Snowball Earth events \citep{Pierrehumbert2005} may have inhibited melting for long enough for hydrohalite crusts to form over wide areas \citep{Light2009}. In regions of net sublimation in the tropics, where low-latitude ice was present and receiving the majority of the solar insolation, climate would have been far more sensitive to sea ice albedo than in higher-latitude ice-covered regions \citep{Light2009}. The high visible and near-IR albedos of hydrohalite crusts, compared with the salt-free ice albedos previously used in climate modeling of low-latitude glacial episodes, could have significantly altered the surface energy balance of Snowball Earth \citep{Carns2016}.

M-dwarf stars are the most common type of stars in the galaxy, and discovering habitable planets around these stars will be the focus of extrasolar planet observational efforts for the foreseeable future, by both large-aperture (30-m class) telescopes on the ground, and space-based observatories such as the Transiting Exoplanet Survey Satellite, (TESS, \citealp{Ricker2014}). Identifying those potentially habitable planets to target for follow-up by the James Webb Space Telescope (JWST, \citealp{Gardner2006}) and the next generation of space missions depends on understanding the impact on habitability of the different interactions between the M-dwarf SED and specific surface types that may be possible on M-dwarf planets, as these interactions defy existing modeling prescriptions.

%. The discovery of a potentially habitable planet orbiting an M-dwarf star only 4.2 light years away (Proxima Centauri b; \citealp{Anglada-Escude2016}) puts surface albedo into even sharper focus as a parameter that could shape the habitability and observability of such planets in the JWST era \citep{Turbet2016, Kreidberg2016}. As such, the interactions between planets' potential surface types and the SEDs of M-dwarf host stars must be understood
Near the outer edge of the habitable zone, where a number of potentially habitable M-dwarf planets have been discovered \citep{Quintana2014, Dittmann2017, Gillon2017}, surface temperatures on regions of a planet may reach well below -23$^\circ$C (depending on atmospheric greenhouse gas concentrations), where hydrohalite formation may occur, making the climatic effect of this possible surface type on M-dwarf planets particularly important to quantify.

%Additionally, the low potential water inventory on M-dwarf planets due to low planetary disk mass \citep{Raymond2007} or ocean evaporation as the result of an extended pre-main sequence host star phase \citep{Luger2015b} may provide the requisite environment of net sublimation for the maintenance of hydrohalite on ice surfaces
In this work we calculate the effect of hydrohalite-induced salt-albedo feedback on the climate of ocean-covered planets orbiting M-dwarf stars, and compare these results to those for similar planets modeled without a parameterization for hydrohalite formation. We identify climate regimes where this surface type could be most relevant and impactful for the surface habitability of M-dwarf planets with Earth-like atmospheres, and discuss differences in the climatic effect of this surface type for planets orbiting brighter, Sun-like stars. 

In Section 2 we present and explain our methods and models used. In Sections 3 and 4 we present and discuss the results and significance of our simulations. In Section 5 we offer concluding remarks and implications of this work for future studies of the potential climates of recently-discovered potentially habitable exoplanets.

\section{Methods and Models} \label{sec:models}
We used the Community Climate System Model (CCSM), a three-dimensional (3D) global climate model (GCM) developed to simulate past and present climate states on the Earth \citep{Gent2011}. With an atmospheric component (The Community Atmosphere Model version 4, or CAM4), the Los Alamos sea ice model (CICE version 4; \citeyear{Hunke2008}), and a 50-meter deep, slab ocean, we refer collectively to this suite of coupled model components as CCSM4, as done in previous work (see e.g., \citealp{Bitz2012, Shields2013, Shields2014, Shields2016a}). The ocean is treated as static, but fully mixed with depth. Simulations that include a fully dynamic ocean are too computationally expensive to permit the exploration of a broad range of forcing parameters as we do in this work. The horizontal angular resolution is 2$^\circ$. We simulated the climates of M- and G-dwarf aqua planets (no land), with circular orbits, Earth's radius, mass, and obliquity, atmospheres with 1-bar surface pressure, and Earth-like levels of CO$_2$.  We simulated both Earth-like (24-hr) and synchronous rotation periods. Atmospheric water vapor was allowed to vary during each simulation according to evaporation and precipitation processes on the surface and in the atmosphere. 

The sea-ice albedo parameterization in CCSM3, as used in Shields \emph{et al.} (\citeyear{Shields2013, Shields2014, Shields2016a}), divides the surface albedo into two bands, visible ($\lambda \leqslant$ 0.7 $\mu$m) and near-IR ($\lambda >$ 0.7 $\mu$m), because it is easier to control than the multiple-scattering scheme in later code versions. The default near-IR and visible band albedos are 0.3 and 0.67 for cold bare ice, and 0.68 and 0.8 for cold dry snow, respectively. We used these default values for our G-dwarf planet climate simulations, as they are tuned for an incident solar spectrum, and yield modern-day Earth climates for simulations of planets receiving 100\% of the modern solar constant. However, for our simulations of M-dwarf planet climates, we calculated the two-band albedos weighted by the specific spectrum of our M-dwarf host star, and used those values (Table 1) for greater accuracy. 

\linespread{1.0}
\begin{table}[!htp] 
\caption{Two-band albedos employed for different temperature regimes (given in degrees Celsius) reached in the GCM, weighted by the spectrum for G-dwarf star the Sun and M-dwarf star AD Leo. For temperatures below $-23^\circ$C, the albedos given below incorporate our hydrohalite parameterization applied to the model. $E$ and $P$ denote water evaporation and precipitation, respectively.} 
\vspace{2 mm}
\centering \begin{tabular}{c c c c c} 
\hline\hline 
Host star & $0^\circ>T>-23^\circ$ & $-23^\circ>T>-40^\circ$ & $T<-40^\circ$ & $E-P<0$ \\  [0.5ex] 
\hline
Band & NIR/VIS & NIR/VIS & NIR/VIS & NIR/VIS\\
M-dwarf & 0.18/0.69 & 0.21/0.80 & 0.88/0.95 & 0.49/0.97 \\ 
G-dwarf & 0.30/0.67 & 0.31/0.84 & 0.90/0.96 & 0.68/0.80 \\[1ex]
\hline 
\end{tabular} 
\label{table:nonlin} 
\end{table}
\clearpage
\pagebreak

Modifications were made to the ice thermal code in CICE4, based on the original model written by Bitz \emph{et al.} (\citeyear{Bitz2001}), to incorporate the bare sea ice albedo change due to crystallization of hydrohalite at low temperatures, and the subsequent formation of a hydrohalite crust. In the model, sea ice is allowed to form as surface temperatures reach the freezing point of liquid water.  Areas of net water precipitation were assigned two-band albedos for salt-free snow with 100-$\mu$m sized grains. For temperatures between freezing and the temperature where hydrohalite begins to precipitate in sea ice (T$<$ -23$^\circ$C), we used two-band albedos for salt-free, ``warm" bare ice; below -23$^\circ$C, we used two-band albedos for cold bare ice with precipitated hydrohalite; and below -40$^\circ$C, we used two-band albedos for a fully-formed hydrohalite crust. Two-band albedos weighted by the spectrum of each host star and used for each temperature regime are given in Table 1.

To allow for the simulation of planets orbiting stars with different SEDs, the percentages of incoming stellar flux in each of the twelve wavelength bands that are input to CAM4 were varied according to the SED of the host star (see e.g., \citealp{Shields2013, Shields2014, Shields2016a}). We employed the solar spectrum obtained from Chance and Kurucz (\citeyear{Chance2010}), and a spectrum for M3V star AD Leo \citep{Reid1995, Segura2005} obtained from the Virtual Planetary Laboratory \footnote{http://vpl.astro.washington.edu/spectra/stellar/mstar.htm}. Flux from the M-dwarf star outside of the CAM4 wavebands for shortwave (incoming) flux were folded into the shortest and longest wavebands so that the entire stellar spectrum was incorporated, as done in prior work (\citealp{Shields2013, Shields2014}).

Although the crystallization of hydrohalite has been shown to start once temperatures reach -23$^\circ$C \citep{Carns2015}, the formation of a hydrohalite crust depends on temperatures staying below this value throughout a diurnal cycle. It is possible that temperatures could rise to a level sufficient for melting of a hydrohalite crust during the simulation. The albedo of a dissolving hydrohalite crust is significantly reduced at all wavelengths compared with fully formed hydrohalite \citep{Carns2016}. Abbot \emph{et al.} (\citeyear{Abbot2010b}) identified a diurnal temperature variation in glaciated regions of $\sim$$\pm$10$^\circ$. We therefore used -40$^\circ$C as the surface temperature requirement for hydrohalite crust formation, to provide a conservative 17$^\circ$ buffer that increased the likelihood that temperatures remained cold enough for application of the hydrohalite crust albedos. 

The climatic effects of a hydrohalite crust are governed by the exposure of an icy planetary surface to incident stellar radiation, and therefore depend on water evaporation exceeding precipitation ($E-P>0$) in the surrounding ice. Where $E-P<0$, it can be expected that snow will cover the planetary surface in that particular location, masking any albedo differences between the different parameterizations in our model. We discuss the effect of regions of net water precipitation on the climatic impact of hydrohalite formation in later sections. 

We ran GCM simulations in which we included the hydrohalite parameterization described above (``HH"), as well as with the default ice albedo parameterization, where bare sea ice near-IR and visible albedos remain fixed at salt-free values for all surface temperatures below freezing (``Control"). We simulated a range of levels of incident stellar insolation (``instellation"), and focus our study here on instellation levels that bracket important climate regimes at fixed, Earth-like levels of CO$_2$ (400 ppmv). 

First, we explore the degree of influence on climate of applying the albedo effects of hydrohalite crust formation in simulations of planets receiving the requisite amount of instellation to yield global mean surface temperatures similar to modern-day Earth. We then compare our results with those for climates where a narrow region of open water is present in the tropics\textemdash so called ``waterbelt" states \citep{Wolf2017a}. Finally, we examine simulations of M-dwarf planets in fully glaciated, ``snowball" states \citep{Kirschvink1992}. Given that M-dwarf habitable-zone planets may \citep{Dole1964, Kasting1993, Joshi1997, Edson2011} or may not \citep{Leconte2015} be synchronously rotating, we simulate planets with both 24-hr and synchronous rotation periods, to bracket the possible effects of rotation period on climate sensitivity to salt-albedo feedback. We also compare the climatic impact of the salt-albedo feedback mechanism induced by hydrohalite formation on M-dwarf planets with that on G-dwarf planets, using select simulations run of G-dwarf aqua planets at similar instellation values. We present our results and comparison in the following section. 
\pagebreak
\section{Results} \label{sec:res}
Figure 2 shows annually-averaged values as a function of latitude for surface temperature, ice cover fraction, surface albedo, and water evaporation minus precipitation ($E-P$), for an M-dwarf planet with a 24-hr rotation period receiving 90\% of the modern solar constant from its host star. It includes both the Control case, without hydrohalite albedos assigned for relevant temperatures, and the HH case, with the temperature-dependent hydrohalite albedo parameterization included. The global mean surface temperature is $\sim$290 K, slightly warmer than modern-day Earth (288 K).

\begin{figure}[!htb]
\centering
\includegraphics[scale=0.27]{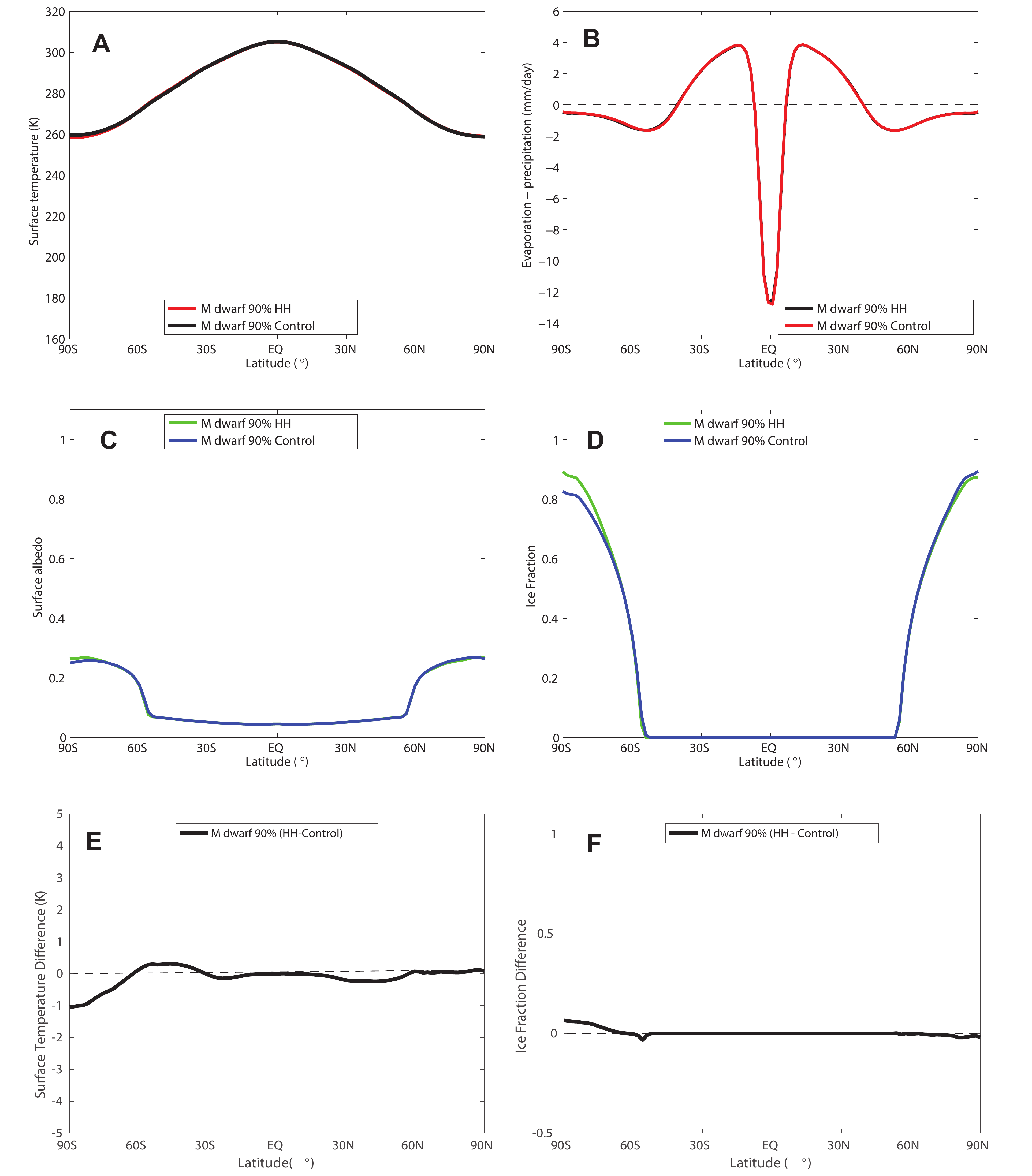}\\
\caption{Annually averaged surface temperature (top left), water evaporation minus precipitation (top right), surface albedo (middle left), ice fraction (middle right), surface temperature difference (bottom left), and ice fraction difference (bottom right) as a function of latitude for an M-dwarf planet with a 24-hr rotation period receiving 90\% of the modern solar constant from its star after 80-year CCSM4 GCM simulations.}
\label{Figure 2.}
\end{figure}

As shown in Figure 2a, surface temperatures do not reach below $\sim$258 K anywhere on the planet in either parameterization at this instellation. The hydrohalite parameterization, which is set in our model to apply higher bare ice albedos when temperatures reach -23$^\circ$C (250 K), is therefore not relevant in this regime, resulting in equivalent patterns of  of $E-P$ on both planets (Figure 2b), and similar surface albedos (Figure 2c). The regions of net evaporation between 0$^\circ$ and $\sim$40$^\circ$S  and 0$^\circ$ and 40$^\circ$N occur where there is no ice present to be exposed to incoming stellar radiation (Figure 2d). We confirm similar results in our simulation of a G-dwarf aqua planet receiving 100\% of the modern solar constant, whose surface climate has been shown in previous work to be analogous to that of an M-dwarf aqua planet at 90\% instellation, due to increased IR absorption by surface water ice and snow, as well as atmospheric CO$_2$ and water vapor \citep{Shields2013}. Differences in surface temperature and ice fraction are minimal (Figures 2e and 2f).

Figure 3 shows similar calculated values for the M-dwarf planet receiving 65\% instellation from its host star for both the HH and Control cases. Our simulations showed this instellation to be the lowest received by the planet without becoming fully glaciated. Here we see that minimum surface temperatures reached $\sim$202 K in both simulations (Figure 3a). There are still largely similar regions of net water precipitation between the two planets everywhere except in the equatorial regions. Here, the water precipitation rate drops slightly at the equator in the HH case compared to the control case (Figure 3b), due to lower amounts of cloud cover on the planet with the hydrohalite parameterization, weaker Hadley circulation, and more shortwave radiation absorbed by the surface at the equator compared to the Control case. Surface albedos (Figure 3c) are slightly higher in the HH case, and global mean surface temperatures begin to differ slightly between the HH and Control cases, at $\sim$249 K and $\sim$ 251 K, respectively, though both planets still have similarly narrow swaths of open water at the equator (Figure 3d). Differences in surface temperature are larger at this instellation (Figures 2e), particularly where albedo differences are larger between the HH and Control cases in the mid to upper latitudes. Ice fraction difference also coincides with net positive $E-P$.

\begin{figure}[!htb]
\centering
\includegraphics[scale=0.27]{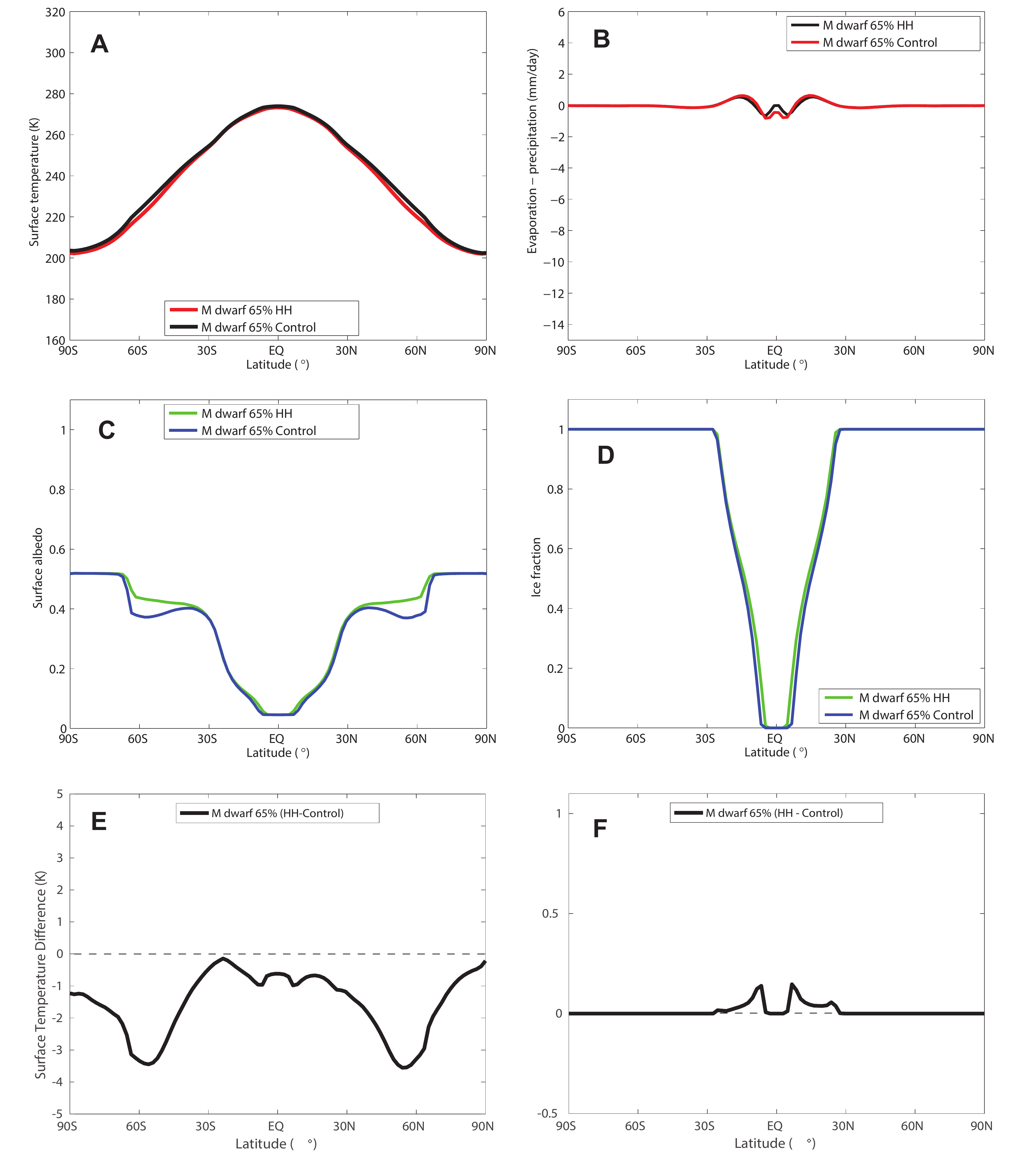}\\
\caption{Annually averaged surface temperature (top left), water evaporation minus precipitation (top right), surface albedo (middle left), ice fraction (middle right), surface temperature difference (bottom left), and ice fraction difference (bottom right) as a function of latitude for an M-dwarf planet receiving 65\% of the modern solar constant from its star after 80-year CCSM4 GCM simulations.}
\label{Figure 3.}
\end{figure}

As M-dwarf planets orbiting in their host stars' habitable zones may be synchronously rotating \citep{Kasting1993, Shields2016b}, we also ran select simulations of planets in such an orbital configuration, with our planets at zero eccentricity, and an obliquity of 23$^\circ$. Figure 4 shows surface temperatures on synchronously-rotating M-dwarf planets at 90\% and 65\% instellation in the Control case (no hydrohalite albedo parameterization), along with ice cover fraction for synchronously-rotating M-dwarf planets at 65\% planets in the HH and Control case. Also shown are $E-P$ and latitudinal change in surface temperature at a range of instellations for planets with Earth-like rotation periods orbiting M- and G-dwarf stars. While the difference between the Control and HH cases is negligible at 90\% instellation for our simulated M-dwarf planets with Earth's rotation period (Figure 2), the colder temperatures exhibited on the day side of the synchronously-rotating M-dwarf planets at equivalent instellation reached $\sim$219 K in the Control case (Figure 4a)\textemdash the regime where hydrohalite could begin to precipitate in sea ice. However, global mean surface temperatures were still approximately equal ($\sim$241 K) in both the Control and HH cases, with similar ice cover fraction. 

\begin{figure}[!htb]
\centering
\includegraphics[scale=0.27]{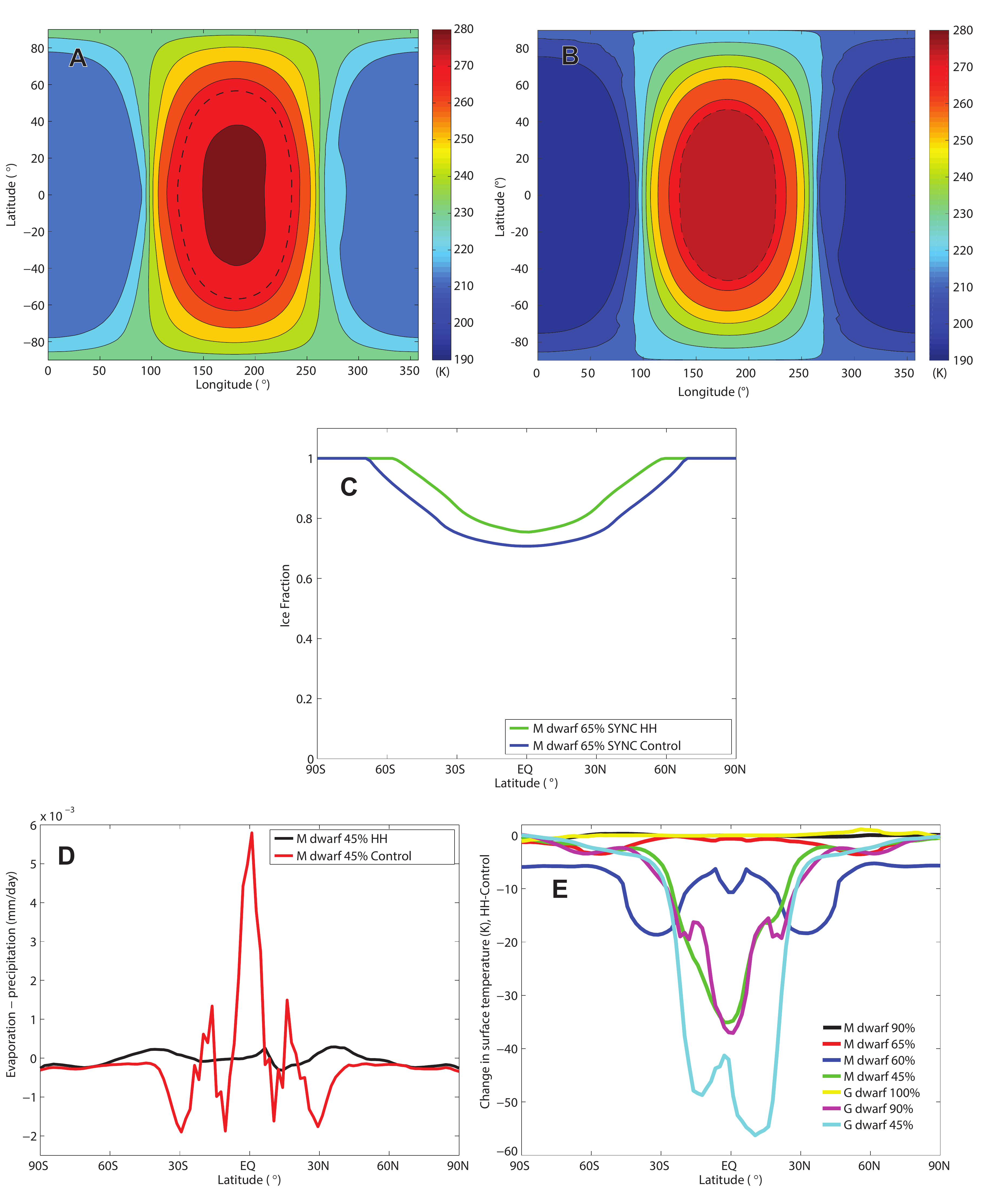}
\caption{Top: Surface temperature as a function of latitude and longitude on synchronously-rotating planets receiving 90\% (left) and 65\% of the modern solar constant from an M-dwarf star (Control case) after 60- and 80-year CCMS4 simulations, respectively. Zero eccentricity and an obliquity of 23$^\circ$ is assumed. The freezing point (273 K) on each planet is labeled by a dashed contour line. Middle: Ice fraction as a function of latitude for a a synchronously-rotating M-dwarf planet receiving 65\% of the modern solar constant from its host star. Bottom left: Water evaporation minus precipitation as a function of latitude for an M-dwarf planet receiving 45\% of the modern solar constant from its star after 80-year CCSM4 simulations. Bottom right: Change in surface temperature as a function of latitude on simulated planets with the hydrohalite parameterization (compared to the Control case) and receiving different levels of instellation from M- and G-dwarf stars after 80 model years of simulation.}
\label{Figure 4.}
\end{figure}

The difference between the HH and Control simulations is stronger in the synchronous case at 65\% instellation, where minimum temperatures reached $\sim$200K on the day side of the planet in the Control case, and $\sim$197 K in the HH case, leading to global mean surface temperatures that were $\sim$10K colder on the planet with the hydrohalite parameterization. This resulted in larger differences in ice cover fraction (Figure 4c) between the two synchronously-rotating cases at 65\% compared to simulations with Earth's rotation period (Figure 3d). 

The differences between the HH and Control cases become even larger at lower instellation. As shown in Figure 4d, although the hydrological cycle has lessened significantly on an ice-covered M-dwarf planet receiving 45\% of the modern solar constant from its host star, the larger relative amount of evaporation in the warmer Control case allows for differences in surface albedo to become much more integral to the resulting climate. Furthermore, the lower the instellation received by the planet from the M-dwarf star, the greater the differences in surface temperature across the planet (Figure 4e), particularly in the tropics where regions of net evaporation are present. The effect of reduced surface temperatures when a hydrohalite parameterization is included is even stronger on G-dwarf planets at low values of instellation, given the increased visible and UV output of their host stars, which has been shown to increase climate sensitivity to water ice-albedo feedback on G-dwarf planets (see e.g., \citealp{Shields2013, Shields2014}). 

\section{Discussion} \label{sec:disc}

Our results indicate that when parameterizations are included in 3D climate models to allow for the possible formation of hydrohalite in bare sea ice on extrasolar planets, the climatic effect of the resulting salt-albedo feedback mechanism could be significant at 65\% and lower values of instellation within the habitable zone, and 90\% and lower values of instellation for synchronously-rotating M-dwarf planets, assuming fixed CO$_2$ levels. The large difference in reflectivity between water ice and hydrohalite in the near-IR make this surface type particularly relevant to consider in simulating planets around M-dwarf stars, as they emit strongly in this region of the spectrum. Such planets could be colder at a given instellation than prior calculations would suggest. 

For M-dwarf planets with global mean surface temperatures similar to modern-day Earth and rotation periods equal to Earth's (24 hours), temperatures do not drop low enough for the crystallization of hydrohalite. However, for synchronously-rotating M-dwarf planets receiving similar instellation, the low temperatures required for hydrohalite crystallization and crust formation (-23$^\circ$C and -40$^\circ$C, respectively) are reached on the day side of these planets (assuming Earth-like atmospheric levels of CO$_2$), even in the middle-range of the habitable zone. This may increase the likelihood of snowball states on synchronously-rotating planets, although active carbon cycles, if present on these planets, may prevent the occurrence of such states \citep{Checlair2017}. 

Orbit precession can cause even those planets affected by tides to retain constant non-zero obliquity, which has been shown to protect a planet from the most severe temperature extremes and improve conditions for surface habitability \citep{Dobrovolskis2009}. Our results with an obliquity of 23$^\circ$ in our simulations of synchronous planets therefore constitute a lower limit on the effects of the hydrohalite parameterization on synchronously-rotating habitable-zone planets compared to those with zero obliquity.  

Though differences in global mean surface temperatures are still negligible between synchronous HH and Control cases in the middle of the habitable zone, we find larger differences in global mean surface temperature at lower levels of instellation within the habitable zone ($\leq$65$\%$ of the modern solar constant), where there is still open water present on the planet, compared to simulations of planets without our hydrohalite parameterization. Synchronous rotation on simulated aqua planets has been shown to cause weakened low-latitude zonal winds \citep{Edson2011} and larger total cloud cover \citep{Yang2014}, which cool the planets compared with planets with faster rotation periods. Hydrohalite may therefore be an important and plausible surface type to form on synchronously-rotating habitable-zone M-dwarf planets such as Proxima Centauri b and TRAPPIST-1e, which receive 65\% and $\sim$60\% of the modern solar constant from their host stars, respectively, and are both expected to be synchronously rotating \citep{Anglada-Escude2016, Gillon2017}. 

Climatic differences between M-dwarf planets with and without a hydrohalite parameterization included in the model are especially large for so-called `waterbelt" planets with narrow swaths of open water in the tropics, and for planets that are colder, assuming an Earth-like amount of CO$_2$. We find that as the instellation is lowered in our simulations, the difference in global mean surface temperature between the planets in the HH and Control cases increases. For our M-dwarf planets receiving 45\% instellation, similar to recently discovered planet LHS 1140b (46\%, \citealp{Dittmann2017}), the global mean surface temperature is 12$^\circ$C colder on the planet with the hydrohalite parameterization. This difference has two important implications. First, planets like LHS 1140b, which is still in the habitable zone of its host star, would likely require more CO$_2$ to build up in their atmospheres, as the silicate weathering rate decreases (e.g., \citealp{Walker1981}), than previously expected to keep the planet warm enough for surface liquid water, as these planets could be much colder than originally assumed. Secondly, frozen planets near or outside of the outer edge of their stars' habitable zones, with only the steady brightening of their host stars over time to depend on to eventually generate above-freezing surface temperatures, would likely have higher instellation thresholds for melting and surface habitability than might otherwise be proposed. Additionally, the resulting cooler climates as a consequence of hydrohalite crystallization may make it more difficult for so-called limit cycles between snowball and warm climates  (e.g., \citealp{Abbot2016, Menou2015, Haaq-Misra2016}) to operate on planets with sufficiently low values of instellation, or introduce limit cycles previously believed unlikely on M-dwarf planets \citep{Haaq-Misra2016}, as a result of hydrohalite-induced salt-albedo-feedback.

Planets orbiting G-dwarf stars exhibited even greater decreases in surface temperature, compared with M-dwarf planets, when the hydrohalite parameterization was included in the simulation. This is due to their increased sensitivity to water ice-albedo feedback, which
contributes significantly to the lower surface temperatures on G-dwarf planets at a given instellation. Our results demonstrate that hydrohalite could have a significant climatic impact on cold, dry G-dwarf planets as well, should they be present at the distant outer regions of their host stars' habitable zones. 

Although we used the default two-band visible and near-IR salt-free ice and snow albedos for our climate simulations of G-dwarf planet climates, we employed values weighted by our G-dwarf spectrum from Chance and Kurucz (\citeyear{Chance2010}) in our albedo parameterization for hydrohalite crystallization and crust formation. The values for salt-free ice and snow calculated with our outside solar spectrum are higher than the default values employed in the GCM, resulting in even colder temperatures on simulated G-dwarf planets compared to our M-dwarf simulations. We therefore elected to include the G-dwarf simulations that used the default salt-free albedos here, because they yield modern-day Earth-like climates at Earth's current instellation level (100\% of the modern solar constant), providing a useful reference point. These simulations therefore constitute a lower limit on the climatic differences possible between G-dwarf planets and M-dwarf planets due to hydrohalite effects.

The climatic effects of hydrohalite presented here depend on the presence of oceans on these planets. Planets composed predominantly of land have been shown to be less susceptible to episodes of global-scale glaciation \citep{Abe2011}, and the small area of sea ice on such planets would limit the effect of salt-albedo feedback. The low potential water inventory on M-dwarf planets, due to low planetary disk mass \citep{Raymond2007} or ocean evaporation as the result of an extended pre-main sequence host star phase \citep{Luger2015b}, could certainly impact the likelihood of hydrohalite formation in sea ice. However, M-dwarf planets subjected to water loss may still retain enough water to maintain habitable surface conditions, depending on their initial water content (e.g., \citealp{Bolmont2017a}).  Additionally, our results would naturally be sensitive to ocean salinity, which is unconstrained on exoplanets. On Earth, salinity levels in young sea ice vary from 12-20 parts per thousand by mass ($^\circ/_{\circ\circ}$) from seawater of normal salinity (32-35 $^\circ/_{\circ\circ}$). The salt is present in the form of brine inclusions; as the sea ice cools, water freezes onto the walls of the inclusions, concentrating the brine until it reaches the temperature and salinity threshold for salt crystallization \citep{Carns2015}. We have adopted a temperature threshold for hydrohalite crystallization appropriate for the mixture of salts present in Earth seawater. On planets with lower seawater salinity, or where sodium chloride is not the most common salt in seawater, it may be more difficult for hydrohalite to form. 

\section{Conclusions} \label{sec:conc}
Using a three-dimensional global climate model modified to incorporate an albedo parameterization for sodium chloride dihydrate, or ``hydrohalite" (NaCl $\cdot$ 2H$_2$O)\textemdash a surface type previously unexplored in the context of extrasolar planets\textemdash we have shown that simulations that include the salt-albedo feedback mechanism generated by the crystallization of this surface type in bare sea ice result in lower planetary global mean surface temperatures compared to simulations without this parameterization included, assuming Earth-like levels of CO$_2$. The large differences between the albedo of hydrohalite and water ice in the near-IR are primarily responsible for the difference in global mean surface temperatures on simulated M-dwarf planets with and without the hydrohalite parameterization included. G-dwarf planets, which exhibit greater climate sensitivity to water ice-albedo feedback and are therefore cooler to begin with at a given instellation, exhibit even stronger planetary cooling in simulations incorporating the albedo effects of this surface type. The climatic effect of hydrohalite becomes particularly important on rapidly-rotating habitable-zone M-dwarf planets receiving 65\% instellation from their host stars, with narrow swaths of open water. Habitable-zone M-dwarf planets that are synchronously rotating and receiving less than 90\% instellation from their host stars exhibit increased susceptibility to the climatic effects of this surface type, due to colder minimum surface temperatures reached on the day side of the planet compared with planets with Earth-like rotation periods receiving similar instellation. The effect is even stronger on fully glaciated planets, where net water evaporation exceeds precipitation such that differences in surface albedo are exposed to incoming stellar radiation, rather than masked by falling snow. These planets are therefore likely to be even colder than originally presumed, given the incorporation of an albedo parameterization for hydrohalite formation in climate simulations. The habitable-zone instellation values where the climatic effects of hydrohalite appear most relevant correspond to those of several recently discovered potentially-habitable M-dwarf planets, including LHS 1140b, TRAPPIST-1e, and Proxima Centauri b; therefore, future simulations of the potential climates of these planets would benefit from inclusion of the hydrohalite albedo parameterization we have developed and incorporated here. By depressing global mean surface temperatures, hydrohalite could increase the greenhouse gas concentration required to maintain surface liquid water on planets near the outer edge of their host stars' habitable zones. It may also increase the instellation values necessary to perpetuate free-thaw cycles on planets with non-zero eccentricities or other sources of climate cycling.

\section{Acknowledgments} \label{sec:acknowl}

This material is based upon work supported by NASA  under grant number  NNH16ZDA001N, which is part of the ``Habitable Worlds" program, by the National Science Foundation under Award Nos. 1401554 and 1753373, by a University of California President's Postdoctoral Fellowship, and a Clare Boothe Luce Professorship. We would like to acknowledge high-performance computing support from Yellowstone (ark:/85065/d7wd3xhc) and Cheyenne (doi:10.5065/D6RX99HX) provided by NCAR's Computational and Information Systems Laboratory, sponsored by the National Science Foundation. A.S. thanks Cecilia Bitz for important discussions and generous help with code modification, Eric Wolf for providing useful code, and John Armstrong, Eric Agol, Shawn Domagal-Goldman, Raymond Pierrehumbert, and Jonathan Mitchell for helpful discussions. We also thank an anonymous reviewer for beneficial feedback that greatly improved the paper.

\newpage
%\bibliography{/Users/aomawashields/Dropbox/FACULTY_POSITION_UCI/RESEARCH/refs}

\begin{thebibliography}{}
\expandafter\ifx\csname natexlab\endcsname\relax\def\natexlab#1{#1}\fi
\providecommand{\url}[1]{\href{#1}{#1}}
\providecommand{\dodoi}[1]{doi:~\href{http://doi.org/#1}{\nolinkurl{#1}}}
\providecommand{\doeprint}[1]{\href{http://ascl.net/#1}{\nolinkurl{http://ascl.net/#1}}}
\providecommand{\doarXiv}[1]{\href{https://arxiv.org/abs/#1}{\nolinkurl{https://arxiv.org/abs/#1}}}

\bibitem[{{Abbot}(2016)}]{Abbot2016}
{Abbot}, D.~S. 2016, \apj, 827, 117, \dodoi{10.3847/0004-637X/827/2/117}

\bibitem[{Abbot {et~al.}(2010)Abbot, Eisenman, \& Pierrehumbert}]{Abbot2010b}
Abbot, D.~S., Eisenman, I., \& Pierrehumbert, R.~T. 2010, Journal of Climate,
  23, 6100, \dodoi{10.1175/2010JCLI3693.1}

\bibitem[{{Abe} {et~al.}(2011){Abe}, {Abe-Ouchi}, {Sleep}, \&
  {Zahnle}}]{Abe2011}
{Abe}, Y., {Abe-Ouchi}, A., {Sleep}, N.~H., \& {Zahnle}, K.~J. 2011,
  Astrobiology, 11, 443, \dodoi{10.1089/ast.2010.0545}

\bibitem[{Anglada-Escud{\'e} {et~al.}(2016)Anglada-Escud{\'e}, Amado, Barnes,
  Berdi{\~n}as, Butler, Coleman, de~la Cueva, Dreizler, Endl, Giesers, Jeffers,
  Jenkins, Jones, Kiraga, K{\"u}rster, L{\'o}pez-Gonz{\'a}lez, Marvin, Morales,
  Morin, Nelson, Ortiz, Ofir, Paardekooper, Reiners, Rodr{\'\i}guez,
  Rodrίguez-L{\'o}pez, Sarmiento, Strachan, Tsapras, Tuomi, \&
  Zechmeister}]{Anglada-Escude2016}
Anglada-Escud{\'e}, G., Amado, P.~J., Barnes, J., {et~al.} 2016, Nature, 536,
  437

\bibitem[{Bitz {et~al.}(2001)Bitz, Holland, Weaver, \& Eby}]{Bitz2001}
Bitz, C.~M., Holland, M.~M., Weaver, A.~J., \& Eby, M. 2001, Journal of
  Geophysical Research: Oceans, 106, 2441, \dodoi{10.1029/1999JC000113}

\bibitem[{{Bitz} {et~al.}(2012){Bitz}, {Shell}, {Gent}, {Bailey},
  {Danabasoglu}, {Armour}, {Holland}, \& {Kiehl}}]{Bitz2012}
{Bitz}, C.~M., {Shell}, K.~M., {Gent}, P.~R., {et~al.} 2012, Journal of
  Climate, 25, 3053, \dodoi{10.1175/JCLI-D-11-00290.1}

\bibitem[{{Bolmont} {et~al.}(2017){Bolmont}, {Selsis}, {Owen}, {Ribas},
  {Raymond}, {Leconte}, \& {Gillon}}]{Bolmont2017a}
{Bolmont}, E., {Selsis}, F., {Owen}, J.~E., {et~al.} 2017, \mnras, 464, 3728,
  \dodoi{10.1093/mnras/stw2578}

\bibitem[{{Brandt} {et~al.}(2005){Brandt}, {Warren}, {Worby}, \&
  {Grenfell}}]{Brandt2005}
{Brandt}, R.~E., {Warren}, S.~G., {Worby}, A.~P., \& {Grenfell}, T.~C. 2005,
  Journal of Climate, 18, 3606

\bibitem[{Carns {et~al.}(2015)Carns, Brandt, \& Warren}]{Carns2015}
Carns, R.~C., Brandt, R.~E., \& Warren, S.~G. 2015, Journal of Geophysical
  Research: Oceans, 120, 7400, \dodoi{10.1002/2015JC011119}

\bibitem[{Carns {et~al.}(2016)Carns, Light, \& Warren}]{Carns2016}
Carns, R.~C., Light, B., \& Warren, S.~G. 2016, Journal of Geophysical
  Research: Oceans, 121, 5217, \dodoi{10.1002/2016JC011804}

\bibitem[{{Chance} \& {Kurucz}(2010)}]{Chance2010}
{Chance}, K., \& {Kurucz}, R.~L. 2010, \jqsrt, 111, 1289,
  \dodoi{10.1016/j.jqsrt.2010.01.036}

\bibitem[{{Checlair} {et~al.}(2017){Checlair}, {Menou}, \&
  {Abbot}}]{Checlair2017}
{Checlair}, J., {Menou}, K., \& {Abbot}, D.~S. 2017, \apj, 845, 132,
  \dodoi{10.3847/1538-4357/aa80e1}

\bibitem[{{Dittmann} {et~al.}(2017){Dittmann}, {Irwin}, {Charbonneau},
  {Bonfils}, {Astudillo-Defru}, {Haywood}, {Berta-Thompson}, {Newton},
  {Rodriguez}, {Winters}, {Tan}, {Almenara}, {Bouchy}, {Delfosse}, {Forveille},
  {Lovis}, {Murgas}, {Pepe}, {Santos}, {Udry}, {W{\"u}nsche}, {Esquerdo},
  {Latham}, \& {Dressing}}]{Dittmann2017}
{Dittmann}, J.~A., {Irwin}, J.~M., {Charbonneau}, D., {et~al.} 2017, \nat, 544,
  333, \dodoi{10.1038/nature22055}

\bibitem[{{Dobrovolskis}(2009)}]{Dobrovolskis2009}
{Dobrovolskis}, A.~R. 2009, \icarus, 204, 1,
  \dodoi{10.1016/j.icarus.2009.06.007}

\bibitem[{{Dole}(1964)}]{Dole1964}
{Dole}, S.~H. 1964, {Habitable Planets for Man} (Blaisdell, New York.)

\bibitem[{{Edson} {et~al.}(2011){Edson}, {Lee}, {Bannon}, {Kasting}, \&
  {Pollard}}]{Edson2011}
{Edson}, A., {Lee}, S., {Bannon}, P., {Kasting}, J.~F., \& {Pollard}, D. 2011,
  \icarus, 212, 1, \dodoi{10.1016/j.icarus.2010.11.023}

\bibitem[{{Gardner} {et~al.}(2006){Gardner}, {Mather}, {Clampin}, {Doyon},
  {Greenhouse}, {Hammel}, {Hutchings}, {Jakobsen}, {Lilly}, {Long}, {Lunine},
  {McCaughrean}, {Mountain}, {Nella}, {Rieke}, {Rieke}, {Rix}, {Smith},
  {Sonneborn}, {Stiavelli}, {Stockman}, {Windhorst}, \& {Wright}}]{Gardner2006}
{Gardner}, J.~P., {Mather}, J.~C., {Clampin}, M., {et~al.} 2006, \ssr, 123,
  485, \dodoi{10.1007/s11214-006-8315-7}

\bibitem[{{Gent} {et~al.}(2011){Gent}, {Danabasoglu}, {Donner}, {Holland},
  {Hunke}, {Jayne}, {Lawrence}, {Neale}, {Rasch}, {Vertenstein}, {Worley},
  {Yang}, \& {Zhang}}]{Gent2011}
{Gent}, P.~R., {Danabasoglu}, G., {Donner}, L.~J., {et~al.} 2011, Journal of
  Climate, 24, 4973, \dodoi{10.1175/2011JCLI4083.1}

\bibitem[{{Gillon} {et~al.}(2017){Gillon}, {Triaud}, {Demory}, {Jehin}, {Agol},
  {Deck}, {Lederer}, {de Wit}, {Burdanov}, {Ingalls}, {Bolmont}, {Leconte},
  {Raymond}, {Selsis}, {Turbet}, {Barkaoui}, {Burgasser}, {Burleigh}, {Carey},
  {Chaushev}, {Copperwheat}, {Delrez}, {Fernandes}, {Holdsworth}, {Kotze}, {Van
  Grootel}, {Almleaky}, {Benkhaldoun}, {Magain}, \& {Queloz}}]{Gillon2017}
{Gillon}, M., {Triaud}, A.~H.~M.~J., {Demory}, B.-O., {et~al.} 2017, ArXiv
  e-prints.
\newblock \doarXiv{1703.01424}

\bibitem[{{Haqq-Misra} {et~al.}(2016){Haqq-Misra}, {Kopparapu}, {Batalha},
  {Harman}, \& {Kasting}}]{Haaq-Misra2016}
{Haqq-Misra}, J., {Kopparapu}, R.~K., {Batalha}, N.~E., {Harman}, C.~E., \&
  {Kasting}, J.~F. 2016, \apj, 827, 120, \dodoi{10.3847/0004-637X/827/2/120}

\bibitem[{Hudson {et~al.}(2006)Hudson, Warren, Brandt, Grenfell, \&
  Six}]{Hudson2006}
Hudson, S.~R., Warren, S.~G., Brandt, R.~E., Grenfell, T.~C., \& Six, D. 2006,
  Journal of Geophysical Research: Atmospheres, 111, n/a,
  \dodoi{10.1029/2006JD007290}

\bibitem[{{Hunke} \& {Lipscomb}(2008)}]{Hunke2008}
{Hunke}, E.~C., \& {Lipscomb}, W.~H. 2008, {CICE: The Los Alamos Sea Ice Model.
  Documentation and Software User's Manual. Version 4.0.} (T-3 Fluid Dynamics
  Group, Los Alamos National Laboratory, Tech. Rep. LA-CC-06-012.)

\bibitem[{{Joshi} \& {Haberle}(2012)}]{Joshi2012}
{Joshi}, M.~M., \& {Haberle}, R.~M. 2012, Astrobiology, 12, 3,
  \dodoi{10.1089/ast.2011.0668}

\bibitem[{{Joshi} {et~al.}(1997){Joshi}, {Haberle}, \& {Reynolds}}]{Joshi1997}
{Joshi}, M.~M., {Haberle}, R.~M., \& {Reynolds}, R.~T. 1997, \icarus, 129, 450,
  \dodoi{10.1006/icar.1997.5793}

\bibitem[{{Kasting} {et~al.}(1993){Kasting}, {Whitmire}, \&
  {Reynolds}}]{Kasting1993}
{Kasting}, J.~F., {Whitmire}, D.~P., \& {Reynolds}, R.~T. 1993, \icarus, 101,
  108, \dodoi{10.1006/icar.1993.1010}

\bibitem[{{Kirschvink}(1992)}]{Kirschvink1992}
{Kirschvink}, J. 1992, {Late Proterozoic Low-Latitude Global Glaciation: the
  Snowball Earth}, ed. J.~Schopf, Vol. The Proterozoic Biosphere: A
  Multidisciplinary Study (Cambridge University Press), 51--52

\bibitem[{{Leconte} {et~al.}(2015){Leconte}, {Wu}, {Menou}, \&
  {Murray}}]{Leconte2015}
{Leconte}, J., {Wu}, H., {Menou}, K., \& {Murray}, N. 2015, Science, 347, 632,
  \dodoi{10.1126/science.1258686}

\bibitem[{Light {et~al.}(2009)Light, Brandt, \& Warren}]{Light2009}
Light, B., Brandt, R.~E., \& Warren, S.~G. 2009, Journal of Geophysical
  Research: Oceans, 114, n/a, \dodoi{10.1029/2008JC005211}

\bibitem[{Light {et~al.}(2016)Light, Carns, \& Warren}]{Light2016}
Light, B., Carns, R.~C., \& Warren, S.~G. 2016, Journal of Geophysical
  Research: Oceans, 121, 4966, \dodoi{10.1002/2016JC011803}

\bibitem[{{Luger} \& {Barnes}(2015)}]{Luger2015b}
{Luger}, R., \& {Barnes}, R. 2015, Astrobiology, 15, 119,
  \dodoi{10.1089/ast.2014.1231}

\bibitem[{{Menou}(2015)}]{Menou2015}
{Menou}, K. 2015, Earth and Planetary Science Letters, 429, 20,
  \dodoi{10.1016/j.epsl.2015.07.046}

\bibitem[{{Pierrehumbert}(2005)}]{Pierrehumbert2005}
{Pierrehumbert}, R. 2005, \jgra, 110, 2156, \dodoi{10.1029/2004JD005162}

\bibitem[{Quintana {et~al.}(2014)Quintana, Barclay, Raymond, Rowe, Bolmont,
  Caldwell, Howell, Kane, Huber, Crepp, Lissauer, Ciardi, Coughlin, Everett,
  Henze, Horch, Isaacson, Ford, Adams, Still, Hunter, Quarles, \&
  Selsis}]{Quintana2014}
Quintana, E.~V., Barclay, T., Raymond, S.~N., {et~al.} 2014, Science, 344, 277,
  \dodoi{10.1126/science.1249403}

\bibitem[{{Raymond} {et~al.}(2007){Raymond}, {Scalo}, \&
  {Meadows}}]{Raymond2007}
{Raymond}, S.~N., {Scalo}, J., \& {Meadows}, V.~S. 2007, \apj, 669, 606,
  \dodoi{10.1086/521587}

\bibitem[{{Reid} {et~al.}(1995){Reid}, {Hawley}, \& {Gizis}}]{Reid1995}
{Reid}, I.~N., {Hawley}, S.~L., \& {Gizis}, J.~E. 1995, \aj, 110, 1838,
  \dodoi{10.1086/117655}

\bibitem[{{Ricker} {et~al.}(2014){Ricker}, {Winn}, {Vanderspek}, {Latham},
  {Bakos}, {Bean}, {Berta-Thompson}, {Brown}, {Buchhave}, {Butler}, {Butler},
  {Chaplin}, {Charbonneau}, {Christensen-Dalsgaard}, {Clampin}, {Deming},
  {Doty}, {De Lee}, {Dressing}, {Dunham}, {Endl}, {Fressin}, {Ge}, {Henning},
  {Holman}, {Howard}, {Ida}, {Jenkins}, {Jernigan}, {Johnson}, {Kaltenegger},
  {Kawai}, {Kjeldsen}, {Laughlin}, {Levine}, {Lin}, {Lissauer}, {MacQueen},
  {Marcy}, {McCullough}, {Morton}, {Narita}, {Paegert}, {Palle}, {Pepe},
  {Pepper}, {Quirrenbach}, {Rinehart}, {Sasselov}, {Sato}, {Seager},
  {Sozzetti}, {Stassun}, {Sullivan}, {Szentgyorgyi}, {Torres}, {Udry}, \&
  {Villasenor}}]{Ricker2014}
{Ricker}, G.~R., {Winn}, J.~N., {Vanderspek}, R., {et~al.} 2014, in \procspie,
  Vol. 9143, Space Telescopes and Instrumentation 2014: Optical, Infrared, and
  Millimeter Wave, 914320

\bibitem[{{Seager}(2013)}]{Seager2013}
{Seager}, S. 2013, Science, 340, 577, \dodoi{10.1126/science.1232226}

\bibitem[{{Segura} {et~al.}(2005){Segura}, {Kasting}, {Meadows}, {Cohen},
  {Scalo}, {Crisp}, {Butler}, \& {Tinetti}}]{Segura2005}
{Segura}, A., {Kasting}, J.~F., {Meadows}, V., {et~al.} 2005, Astrobiology, 5,
  706, \dodoi{10.1089/ast.2005.5.706}

\bibitem[{Shields {et~al.}(2016b)Shields, Ballard, \& Johnson}]{Shields2016b}
Shields, A.~L., Ballard, S., \& Johnson, J.~A. 2016b, Physics Reports, 663, 1 ,
  \dodoi{http://dx.doi.org/10.1016/j.physrep.2016.10.003}

\bibitem[{Shields {et~al.}(2016a)Shields, Barnes, Agol, Charnay, Bitz, \&
  Meadows}]{Shields2016a}
Shields, A.~L., Barnes, R., Agol, E., {et~al.} 2016a, Astrobiology, 16, 443,
  \dodoi{10.1089/ast.2015.1353}

\bibitem[{{Shields} {et~al.}(2014){Shields}, {Bitz}, {Meadows}, {Joshi}, \&
  {Robinson}}]{Shields2014}
{Shields}, A.~L., {Bitz}, C.~M., {Meadows}, V.~S., {Joshi}, M.~M., \&
  {Robinson}, T.~D. 2014, \apjl, 785, L9, \dodoi{10.1088/2041-8205/785/1/L9}

\bibitem[{{Shields} {et~al.}(2013){Shields}, {Meadows}, {Bitz},
  {Pierrehumbert}, {Joshi}, \& {Robinson}}]{Shields2013}
{Shields}, A.~L., {Meadows}, V.~S., {Bitz}, C.~M., {et~al.} 2013, Astrobiology,
  13, 715, \dodoi{10.1089/ast.2012.0961}

\bibitem[{{Walker} {et~al.}(1981){Walker}, {Hays}, \& {Kasting}}]{Walker1981}
{Walker}, J.~C.~G., {Hays}, P.~B., \& {Kasting}, J.~F. 1981, \jgr, 86, 9776,
  \dodoi{10.1029/JC086iC10p09776}

\bibitem[{{Wolf} {et~al.}(2017){Wolf}, {Shields}, {Kopparapu}, {Haqq-Misra}, \&
  {Toon}}]{Wolf2017a}
{Wolf}, E.~T., {Shields}, A.~L., {Kopparapu}, R.~K., {Haqq-Misra}, J., \&
  {Toon}, O.~B. 2017, ArXiv e-prints.
\newblock \doarXiv{1702.03315}

\bibitem[{{Yang} {et~al.}(2014){Yang}, {Bou{\'e}}, {Fabrycky}, \&
  {Abbot}}]{Yang2014}
{Yang}, J., {Bou{\'e}}, G., {Fabrycky}, D.~C., \& {Abbot}, D.~S. 2014, \apjl,
  787, L2, \dodoi{10.1088/2041-8205/787/1/L2}

\end{thebibliography}

\end{document}